\documentstyle[11pt,aaspp4,flushrt]{article}
\def \beq{\begin{equation}}
\def \eeq{\end{equation}}

\def \bh{\rm bh}
\def \cl{\rm cl}
\def \SI{\rm SIS}

\def \apj{ApJ}
\def \deg{\hbox{$^\circ$}}
\def \ldash{\mbox{---}}

\begin{document}

\title{A reassessment of the data and models of the gravitational 
lens Q0957+561}

\author{Rennan Barkana\footnote{email: barkana@ias.edu}}
\affil{Institute for Advanced Study, Olden Lane, Princeton,
NJ 08540, USA}

\author{Joseph Leh\'ar, Emilio E. Falco, Norman A. Grogin, Charles
R. Keeton, Irwin I. Shapiro}
\affil{Center for Astrophysics, 60 Garden St, Cambridge, MA
02138, USA}

\begin{abstract}
We examine models of the mass distribution for the first known
case of gravitational lensing. Several new sets of constraints 
are used, based on recent observations. We remodel the VLBI 
observations of the radio cores and jets in the two images of Q0957+561, 
showing that the previously derived positions and uncertainties 
were incorrect. We use as additional constraints the 
candidate lensed pairs of galaxies discovered recently with the 
Hubble Space Telescope. We explore a wider range of lens models 
than before, and find that the Hubble constant is not tightly 
constrained once elliptical lens models are considered. 
We also discuss the systematic uncertainties caused by the 
cluster containing the lens galaxy. We conclude that additional 
observations of the candidate lensed galaxies as well as direct 
measurements of the cluster mass profile are needed before 
a useful value of the Hubble constant can be derived from
this lens.
\end{abstract}


\vspace{.1in}

\section{Introduction}

Research on gravitational lensing has grown substantially during the
past two decades (see,~e.g., Schneider et~al.\ 1992; Blandford \& Narayan 1992).
A major reason for this attention is the prospect of obtaining an
estimate of Hubble's constant, $H_0$, directly from cosmologically
distant sources (Refsdal 1964, 1966), bypassing the many 
calibration-sensitive rungs of the cosmic distance ladder. Such an estimate
requires the measurement of a difference $\Delta\tau$ between 
the arrival times of light from a source via two image paths,
and an accurate model of the lensing mass distribution.
There are now five lensing systems 
for which estimates of $H_0$ have been published:
0957+561; PG1115+080 (Schechter et~al.\ 1997; Barkana 1997;
Impey et~al.\ 1998);
B0218+357 (Biggs et~al.\ 1999); CLASS1608+656 
(Fassnacht et~al.\ 1997); and HE~1104--1805 (Wisotzki et~al.\ 1998).

The ``double quasar'' Q0957+561 provided the first documented case of
gravitational lensing (Walsh et~al.\ 1979). This system includes
two images, $A$ and $B$,
separated by $\sim 6\arcsec$ on the sky, of a
single background quasar at $z$=1.41 and a lensing galaxy $G1$, at $z$=0.36\,.
Monitoring started almost immediately 
after discovery, with the goal of measuring $\Delta\tau$. 
However, it was found to be a very challenging measurement to make. 
Optical (Lloyd 1981; Keel 1982; Florentin-Nielsen
1984; Schild \& Cholfin 1986; Vanderriest et~al.\ 1989; Schild \& Thomson
1995) and radio (Leh\'ar et~al.\ 1992) monitoring programs produced 
extensive data, but analyses with a host of sophisticated techniques 
(see,~e.g., Press, Rybicki \& Hewitt 1992a, 1992b; Pelt et~al.\ 1994, 1996)
could not resolve the conflict between groups obtaining delays near
400 days and those finding delays close to 540 days.
Only recently has an optical detection of a sharp event in 
the light curve of each image resulted in a precise determination
($\Delta\tau=417\pm3$\,days, Kundi\'{c} et~al.\ 1997),
confirming the short value of the delay first obtained by
Schild \& Cholfin (1986).
Additional confidence in this measurement comes from the consistency
with the latest results from radio monitoring (Haarsma et~al.\ 1998).

The other essential ingredient for obtaining the value of $H_0$
is a well-constrained model for the lensing mass distribution. 
A major complication for Q0957+561 is that a cluster of galaxies
surrounding $G1$ also contributes to the lensing (Young et~al.\ 1981).
Lensing mass can be exchanged in models between the cluster and $G1$
without affecting the image configuration, but yielding different
$\Delta\tau$ predictions.  This {\it cluster degeneracy}
is an example of the mass-sheet degeneracy
in lensing identified by Falco et~al.\ (1985). Thus a direct
measurement of the mass of either the galaxy or the cluster is
required to remove the degeneracy and provide an estimate of $H_0$.
The cluster's mass distribution can be estimated from ``weak lensing'',
the shape distortions of background galaxies due to lensing by the
cluster. Such an estimate has been attempted for the Q0957+561 cluster 
(Fischer et~al.\ 1997). However, this cluster is not very massive, 
and thus the effect is weak and the estimate imprecise. 
The cluster potential can also be estimated from X-ray emissions.
This measurement was made only crudely with ROSAT (Chartas et~al.\ 
1998), but observations with the Advanced X-ray Astrophysics Facility
(AXAF) should yield considerable new information.
As for the lens galaxy, its mass can be estimated from velocity 
dispersion measurements.  
Falco et~al.\ (1997) measured a $G1$ velocity dispersion of
$279\pm12$\,km\,s$^{-1}$, improving on an earlier result by
Rhee (1991). There was, however, a difference of $50\pm20$\,km\,s$^{-1}$ 
between the velocity dispersion measured within or outside a
radius of $0\farcs2$ from the galaxy center. 
In a more recent observation, Tonry \&~Franx (1998) measured
a $G1$ velocity dispersion of $288\pm9$\,km\,s$^{-1}$,
which confirms the Falco et~al.\ result, although they found 
no evidence of a velocity dispersion gradient. 
However, converting from these stellar dispersions to 
mass estimates is fraught with uncertainty.
Romanowsky \& Kochanek (1998) have considered 
combining the velocity dispersion measurement with detailed modeling 
of the stellar velocity distribution in the lens galaxy $G1$, but
this combination relies on assumptions regarding the velocity
dispersion profile of $G1$.

The most recent effort to explore models of 0957+561 was by 
Grogin \& Narayan (1996, hereafter GN).
They considered two types of models for the lens and approximated the effect 
of the surrounding cluster as an additional constant shear term. The basic
model type explored by GN represents the lens galaxy as a 
softened power-law sphere (SPLS),
a density profile which allows for both a core radius and an arbitrary 
radial power-law index. The other type of model was adopted from
earlier work by Young et~al.\ (1980) and 
Falco et~al.\ (1991, hereafter FGS). In this latter type, the galaxy
has a King profile, a generalization of the singular isothermal sphere which
switches from constant surface density at the center to an isothermal 
profile at large radii. These models are strongly constrained by
VLBI data which resolve each of the two images into 
a core and several jet components
(Gorenstein et~al.\ 1988; Garrett et~al.\ 1994, hereafter G94).
The positions of the core and jet components
were estimated by G94 from the VLBI visibility amplitudes 
and phases; these positions were used to determine a relative
magnification matrix for the two images, with spatial gradients
along the direction of the jet. The magnification matrix and 
gradients were used by GN to constrain lens models. 
One concern in evaluating the results of GN is the poor reduced $\chi^2$
$(\sim 4)$ of even their best-fitting lens models. 

Using an earlier set of VLBI constraints, Kochanek (1991) modeled
the galaxy and the cluster as potentials expanded to quadrupole
order. Kochanek showed that the mass model is not well constrained
by the lensing data alone, and that deriving a precise estimate of 
Hubble's constant depends on interpreting other
observations such as the stellar velocity dispersion. 

In this paper we reconsider the data and the lens models for Q0957+561.
In \S 2 we model the radio core and jet components in the
two images using the VLBI visibility data of G94 as constraints, 
after finding substantial problems in previous work, both 
in estimates of parameter values and in determinations of
standard errors and correlations. 
In \S 3 we summarize the other observations which we use to constrain 
lens models, including the Hubble Space Telescope (HST) 
observations of Bernstein et~al.\ 
(1997, hereafter B97). In \S 4 we discuss the lens models that
we use. We add ellipticity parameters to the spherical models used
by GN and FGS, and we include a cluster in the model.
We show that for a given cluster density profile 
the lens constraints can, in principle,
determine the center of mass and total mass of the cluster.
We present our results in \S 5 and
discuss their implications. Finally, in \S 6 we summarize our present 
understanding of Q0957+561, and consider the prospects
for a useful $H_0$ determination. 

\section{VLBI Constraints}
\label{09vlbi}

While optical maps of the images of Q0957+561 can yield only their
positions, radio VLBI maps have resolved the images 
and revealed internal structures.
Early VLBI observations (Porcas et~al.\ 1981) found that both 
components have a core-jet radio structure. Improved maps (Gorenstein et 
al.\ 1988) resolved the $A$ and $B$~images each into a compact core
with several jet components, 
enabling a reliable determination of the 
relative magnification matrix of the $A$ and $B$ images. The maps of 
G94 further resolved the images into six components each, denoted 
$A_{1 \ldots 6}$ and $B_{1 \ldots 6}$ (where
$A_1$ and $B_1$ denote the cores), and provided sufficient precision
to estimate the spatial gradient of the 
relative magnification matrix. 

G94 modeled the flux distribution of each component as an elliptical
two-dimensional Gaussian, and estimated the parameter
values for each image from separate fits to the VLBI visibility data. 
These component positions cannot be used directly as constraints
on the lens model because their spatial separation is small
compared to the length scale over which the lens magnification changes
appreciably.  Thus, the position constraints from
the six pairs of corresponding components are highly degenerate.
G94 addressed this concern by introducing a second step in their analysis
(following FGS), using the VLBI components to determine
a relative image magnification matrix and its spatial derivatives 
along the jet direction.
GN used this magnification matrix and its derivatives to constrain their lens models.

We have found substantial problems with the VLBI component 
positions and error estimates determined by G94.
First, G94 used the Caltech VLBI package program MODELFIT to determine the
six Gaussian components.  MODELFIT used only single-precision computations,
and stopped searching long before having converged on a least-squares solution.
Second, G94 used the program ERRFIT in the same package to estimate 
parameter variances and covariances.  We found serious mistakes in ERRFIT,
including inconsistencies with MODELFIT, 
and an error which caused ERRFIT to ignore one third of the data.
Third, since $A$ and $B$ are images of a common source,
and since the extended jet components are not expected to vary
on time scales comparable to $\Delta\tau$, 
we expect the flux density ratios of corresponding
components along the jet to vary smoothly with position, 
according to a small macroscopic magnification gradient.
But because G94 used only partial flux density information, with the 
component positions as constraints in their second step,
their jet flux density ratios
deviate strongly from a smooth gradient.

We have corrected the errors in both MODELFIT and ERRFIT,
and have changed the estimation procedure by combining 
the two steps of the G94 analysis into a single step. 
We take two sets of elliptical Gaussian components
(with six flux density, position, and shape parameters for each image),
and restrict some of the $B$~image parameters to correspond
to those of the $A$~image, through a linear magnification matrix
and its spatial derivatives along the jet direction.
We simultaneously fit this combined set of 
image component and magnification matrix parameters to the VLBI visibility
data for the $A$ and $B$ images. 
In contrast to G94, who ignored the important correlations
of the parameter estimates in the second step of their analysis,
our one-step derivation of the magnification matrix and spatial derivatives
fully incorporates the parameter covariances.
We required the total flux densities and center positions
of the $B$~image jet components
to map to those of the corresponding $A$~image jet components, 
through the relative magnification transformation (see Appendix),
but accounted for limitations of our mapping and flux model
by allowing the VLBI component shapes to vary independently.
Since the core flux density varies perceptibly over time scales of years,
we allowed the $B$~image core flux density (i.e., $B_1$) to be 
independent of the $A_1$ flux density.
Thus, the overall fit involved 59 parameters:
36 $A$~image component parameters, minus 2 to fix $A_1$ at the origin;
4 magnification matrix and 2 independent spatial derivative parameters;
18 $B$~image component shape parameters;
and one for the $B_1$ flux density.
The overall fit yielded a reduced chi-squared of 
$\bar{\chi}^2=2.2$,
for 21040 visibility amplitudes and phases.

Although our improved analysis should produce more reliable parameter estimates,
there are a number of reasons to treat the derived uncertainties conservatively.
First, there is an issue of possible time dependence. 
Parsec-scale jet components often move outwards at
apparently superluminal velocities (see,~e.g., Cawthorne 1991). With an
apparent speed of, say, 6 times the speed of light, the jet components in 0957
would move roughly $0.7$~milliarcseconds (mas) in the time $\Delta\tau$.
Since we are comparing VLBI
observations of $A$ and $B$ obtained {\it simultaneously},
while the lens models assume a stationary source,
superluminal motion could affect our conclusions.
Campbell et~al.\ (1995) monitored the inner part of the VLBI jet over 6 years,
and found no motion of the jet components with respect to the core,
although any change in position of $\sim1\,$mas over this period
would have been detected.  This limit is still 
somewhat larger than our estimated
errors for some of the component positions (see below),
and hence we cannot exclude the possibility of 
superluminal motions being relevant.
A~second concern is substructure in the
lens galaxy (see,~e.g., Mao \& Schneider 1998). Either a globular
cluster or a mass fluctuation in the lens of $\sim$10$^6\,M_{\odot}$
along the path of light from a jet component can deflect this component
by about 1~mas.  Such deflections would occur independently
in each image, and would not be modeled by macro-lens 
models that assume smooth density distributions on arcsecond scales.
Finally, although we have followed G94 in using
a flux density model of multiple elliptical components,
this simple model may not represent the actual 
flux density distribution satisfactorily,
thereby resulting in $\bar{\chi}^2>1$ and underestimated uncertainties.
To account for the imperfect fit, we increase our parameter error estimates
by a factor of $\sqrt{\bar{\chi}^2}$.  This increase corresponds to rescaling
the VLBI data uncertainties to make $\bar{\chi}^2=1$.

The VLBI image component parameter values resulting
from our fit are shown in Table~1.
Except for the $B_1$ flux density, 
the $B$~image positions and flux densities are derived from 
their $A$~image counterparts, through
the image magnification matrix and its spatial derivatives.
Each image component is described by a total flux density, 
a center position, the major axis full-width at half-maximum,
the axis ratio, and the position angle of the major axis. 
We give component centers in relative right ascension 
$\Delta\alpha$ and declination $\Delta\delta$, rather 
than in G94's polar coordinates.
Throughout this paper our coordinates refer to epoch B1950.0\,
which was used for the G94 VLBI observations.
The standard errors in Table~1 have been scaled 
to make $\bar{\chi^2}=1$.
We do not show the parameter covariances, since these parameters
were not used as direct constraints for lens models.
Figure~1 shows the flux density contours of the $A$ and $B$ images
as given by our model. Each image component is also represented in 
the figure by a rectangle whose dimensions coincide 
with the major and minor axes of the corresponding Gaussian.
Note that the $B$~image components agree more closely 
with their $A$~image counterparts than is the case for G94's 
VLBI component model (see Figure~3 in G94). 

Table~2 presents the VLBI constraints that we use
in developing lens models \footnote{The complete covariance
matrix for the parameters of the VLBI fit is available
at http://www.sns.ias.edu/$^{\sim}$barkana/0957.html}.
These constraints include the parameter estimates, 
with scaled standard errors and normalized correlation coefficients,
for the magnification matrix at the core and brightest jet component positions,
and separately for the 
positions of the brightest jet components ($A_5$ and $B_5$)
relative to their respective cores.
Although the jet component positions in Table~2 
are not independent of the magnification matrix,
they do provide the most precise information on the jet structure.
Thus we follow GN by including them as direct constraints to the lens models.
Our results imply a relative $A\rightarrow B$ magnification
of $0.74\pm 0.06$ at the core and $0.64\pm 0.04$ at the brightest jet 
component. The gradients of the eigenvalues along the jet direction from
$A_1$ to $A_5$ (see Appendix) are $\dot{M}_1=(-2.6\pm 1.3)
\times10^{-3}\ {\rm mas}^{-1}$
and $\dot{M}_2=(4.0\pm 3.3)\times10^{-4}\ {\rm mas}^{-1}$.
These gradients differ somewhat from the values estimated by G94, 
of $\dot{M}_1=(0.5\pm 1.7)\times 
10^{-3}\ {\rm mas}^{-1}$ and $\dot{M}_2=(2.6\pm 0.9)\times 
10^{-3}\ {\rm mas}^{-1}$.

\section{Other Observational Constraints}
\label{09s3}

In addition to the information provided by the VLBI structures,
there are a number of other observations that we use 
to constrain the lens models.

The separation between the two quasar images determines
the mass scale of the lens model.
For the $A-B$ core separation, FGS and GN adopted the value of 
$(-1\farcs25271,6\farcs04662)$ with $0\farcs00004$ uncertainty from 
the original measurement of Gorenstein et~al.\ (1984). There seems, 
however, to have been a slight error in FGS in the conversion from 
seconds to arcseconds. We use the correct value of
$(-1\farcs25254,6\farcs04662)$. The difference is tiny and has 
a negligible effect on the results. 

The position of the principal lens galaxy, $G1$,
provides an important constraint.
GN assumed the optical center of 
brightness of $G1$ to be at $(0\farcs 19, 1\farcs 00)$ (Stockton 1980) 
from image B, with an uncertainty in each component of 30 mas.
The lens galaxy is also detectable at radio wavelengths,
and the most precise VLBI observations
of the faint radio component $G'$ yielded an estimated 
separation from $B$ of
$(0\farcs 181, 1\farcs 029)$, with a standard error of 1~mas
(Gorenstein et~al.\ 1983). 
Recent HST observations (B97) yield a $G1$ position
of $(0\farcs 1776, 1\farcs 0186)$ with $3.5$ mas errors,
only about three standard deviations away from the VLBI $G'$ position.
It is not certain whether the position of the radio source
or even that of the optical center coincides with the 
center of the lens potential within the measurement uncertainties,
but as a conservative option, we have chosen to use the B97
$G1$ coordinates and errors to constrain the lens position.

In addition to the VLBI constraints G94 included two 
$B/A$ magnification ratios as constraints: those observed
at the core and at the position of the brightest jet component.
Since this jet flux ratio is incorporated in our VLBI fitting,
we take only the core magnification ratio as an additional constraint.
The core flux density varies over times comparable to $\Delta\tau$.
To account for this, we allowed for a variable core flux ratio
in the VLBI fitting, but the jet constraints alone 
yield a predicted magnification ratio at the core. 
The core magnification ratio has been independently 
determined to be $0.747 \pm 0.015$, from a combination of optical 
emission line ratios with VLA and VLBI light 
curve analyses (see,~e.g., Conner et~al.\ 1992).  
If we add the directly
observed core magnification ratio, we have two constraints
on the same quantity and an additional degree of freedom. 

Models with a smooth surface mass density for the lens
produce a third image of 
Q0957+561, typically demagnified and near the center of the lens galaxy.
No such image has been seen down to a $5\sigma$ limit of $1/30$ the
flux density of image~$B$ (Gorenstein et~al.\ 1984). 
We follow the approach of GN 
in penalizing models only when their predictions exceed this $5\sigma$ limit,
which GN achieve by adding to the $\chi^2$ a term
   \beq
   \chi^2_{C/B}= \left\{
   \begin{array}{ll}
   0 & C/B < 1/30 \\
   \frac{(C/B-1/30)^2}{(1/150)^2} & C/B > 1/30
   \end{array} \right\} ,
   \eeq
where $C/B$ refers to the third image flux density ratio with respect
to the $B$~image.
In the SPLS model, the core radius determines the degree of central mass 
concentration and is the parameter most sensitive to the third-image
flux limit. In the FGS model the central point mass prevents a
third image from forming. Following GN, we add this constraint only 
in cases like the SPLS model where the third-image limit plays a role.

B97 discovered a faint arc with two bright ``Knots'' and a number of ``Blobs'';
B97 noted that the Knots, which form part of a single arc,
appear to be images of each other, 
if the arc is indeed produced by gravitational lensing, 
and that two Blobs (2~and~3) are also multiple images
of a background galaxy. These Blobs may differ somewhat in their
peak surface brightness, but this difference may be an artifact of 
limited angular resolution. 

We summarize the various constraints used in our model fitting in Table~3,
which defines our fiducial, ``full'' set of constraints.
Additional global constraints are given by the extended radio lobes 
found with the VLA (components $C$, $D$ and $E$ of Greenfield et~al.\ 1985),
which must not be multiply imaged by a lens
model. We check for this constraint but do not formally include it 
in the $\chi^2$ since our models always satisfy it easily.

\section{Lens Models}
\label{09s4}

For modeling the gravitational lensing of Q0957+561, we consider 
a lens at redshift $z_L$ and a source at $z_S$, with corresponding
angular diameter distances to the observer $D_L$ and $D_S$,
and a lens-source distance $D_{LS}$. For a deflecting
mass localized in a plane perpendicular to the line of sight,
we write the lens equation (see,~e.g., 
Schneider et~al.\ 1992) as
   \beq
   \vec{\beta}=\vec{\theta}-\vec{\alpha}\ ,
   \eeq
where $\vec{\beta}$ is the source position, $\vec{\theta}$ is the
image position, and $\vec{\alpha}$ is the deflection angle 
scaled by $D_{LS}/D_S$, all measured in the lens plane with the
center of mass of the lens at the origin.
We denote the mass density of the lens
projected on this plane by $\Sigma$ and define a critical density
$\Sigma_c=c^2 D_S/(4 \pi G D_L D_{LS})$. Then (in angular units)
$\vec{\alpha}$ is the gradient of the two-dimensional potential
$\psi$ which is determined by 
   \beq \nabla^2\psi=2\frac{\Sigma}
   {\Sigma_c}\equiv 2\kappa\ . \eeq
If we have one lens but multiple sources at different redshifts,
then to determine the corresponding $\vec{\alpha}$, 
a given deflection angle must be scaled by the appropriate factor
of $D_{LS}/D_S$. Therefore, to account for the HST Blob and Knot
sources in Q0957+561, whose distances are unknown, we must add 
additional parameters $f_{\rm blob}$ and $f_{\rm knot}$ 
which are the $D_{LS}/D_S$ ratios 
for each of these sources over the same ratio for the quasar.

Because of their simplicity, axisymmetric mass distributions
are often used to model gravitational lenses.  The Softened
Power-Law Sphere (SPLS) model, defined by GN, 
can account for physical profiles ranging from isothermal to a 
point-mass, with the added possibility of a softened core. 
Most elliptical galaxies have central cusps in their luminosity
profiles (e.g., Gebhardt et~al.\ 1996). The possible existence of
core radii in dark matter halos is unresolved, with some 
simulations finding a shallow inner density profile with a 
large scatter among halos (Kravtsov 
et~al.\ 1998), while others find a density profile steeper than 
$1/r$ (e.g., Moore et~al.\ 1998).
The SPLS model has a spherically symmetric 
volume density profile,
   \beq
   \rho(r)=\rho_0 \left(1+\frac{r^2}{r_c^2}\right)^{(\eta-3)/2},
   \eeq
with a corresponding projected surface density
   \beq
   \Sigma(\xi)=\Sigma_0 \left(1+\frac{\xi^2}{r_c^2}\right)^{(\eta-2)/2}\ ,
   \eeq
where $\Sigma_0=\rho_0 r_C B(1/2,1-\eta/2)$ and $B$ is the Euler beta
function. The deflection law is
   \beq
   \vec{\alpha}(\vec{\theta}\,)=\left(\frac{\alpha_E^2}{\theta^2}\right)
   \left[\frac{(\theta^2+\theta_c^2)^{\eta/2}-\theta_c^{\eta}}
   {\alpha_E^{\eta}}\right] \vec{\theta}\ ,
   \eeq
where $\alpha_E=\alpha_0^{2/(2-\eta)} \theta_C^{-\eta/(2-\eta)}$ and
in radians 
   \beq
   \alpha_0=\left(\frac{8 \pi G \Sigma_0 r_c^2}{c^2 D \eta}\right)^{1/2}\ ,
   \eeq
with $D=D_L D_S /D_{LS}$. We note that the corresponding dimensionless
surface density (i.e.\ convergence) is 
   \beq \kappa(\theta)=\frac{\eta}{2} \alpha_E^{2-\eta}
   (\theta^2+\theta_c^2)^{\frac{\eta}{2}-1}\ . \eeq
The parameters are thus a normalization
$\alpha_E$, core radius $\theta_c$, and power-law index $\eta$.

We also use the empirical model introduced by FGS, which
consists of a King profile and a central point mass. 
FGS adopted an analytic approximation introduced by Young et~al.
(1981) for the deflection law of the King profile:
   \begin{eqnarray}
   \vec{\alpha}(\vec{\theta}\,)\ [{\rm radians}] & 
   = & \left(\frac{D_L}{D}\right)
   \left(\frac{\sigma_v^2}{c^2}\right) \alpha_*(\theta)\ \hat{\theta}\ , \\
   \alpha_*(\theta) & = & 53.2468\, f\left(1.155 \frac{\theta}{\theta_c}\right)
   - 44.0415\, f\left(0.579 \frac{\theta}{\theta_c}\right)\ , \nonumber \\
   f(x) & = & \frac{\sqrt{1+x^2}-1}{x}\ . \nonumber
   \end{eqnarray}
The parameters are a velocity dispersion $\sigma_v$ and a core radius
$\theta_c$. The corresponding convergence is 
   \begin{eqnarray}
   \kappa(\theta) & = & \left(\frac{D_L}{D}\right)\left(\frac{\sigma_v^2}
   {c^2}\right)\left\{
   \frac{30.75}{\theta_c}g\left(1.155 \frac{\theta}{\theta_c}\right)-
   \frac{12.75}{\theta_c}g\left(0.579 \frac{\theta}{\theta_c}\right)
   \right\}\ ,
   \label{kapfgs} \\ g(x) & = & \frac{1}{\sqrt{1+x^2}}\ . \nonumber
   \end{eqnarray}

In order to fit the data, FGS also included a point mass
of mass $M_{\bh}$ at the center of the galaxy, which yields
   \beq
   \vec{\alpha}(\vec{\theta}\,)=\left(\frac{\alpha_{\bh}^2}{\theta^2}\right)
   \vec{\theta}\ ,
   \eeq
where the Einstein radius is
   \beq
   \alpha_{\bh}=\left(\frac{4 G M_{\bh}}{c^2 D}\right)^{1/2}=0\farcs
   91 \left(\frac{M_{\bh}}{10^{11}\, h^{-1}\, M_{\odot}}\right)^{1/2}\ .
   \eeq
Fitted models imply this point mass is $\sim 10^{11}$ $M_{\odot}$,
much larger than expected for black holes, so this
term should be interpreted as correcting the King profile which by
itself is not steep enough near the center of the lens. Of course, the
mass of the point mass may be redistributed in any axisymmetric
manner (inside the $B$ image radius) without affecting the
lensing, so the FGS model is not necessarily unrealistic. This 
ambiguity of the FGS model with respect to the central 
distribution of mass
in the lens galaxy $G1$ makes it difficult to utilize velocity dispersion
measurements to break the $H_0$ degeneracy. 

Since galaxies are usually not observed to be axisymmetric, 
elliptical mass distributions offer more general and realistic lens 
models. They are difficult to use, however, since the deflection angle
obtained by Schramm (1990) for general elliptical models requires the 
evaluation of
rather slow numerical integrals. To add ellipticity to the
lens model while avoiding this difficulty, GN used an elliptical
potential model. The imaging properties of elliptical potentials have 
been investigated extensively (Kovner 1987, Blandford \& Kochanek
1987 and Kochanek \& Blandford 1987). They become identical
to elliptical densities for very small ellipticities and
produce similar image configurations even for moderate
ellipticity (Kassiola \& Kovner 1993). However, elliptical
potentials cannot represent mass distributions with ellipticities
exceeding about $0.5$ because the 
corresponding density contours acquire the artificial 
feature of a dumbbell shape, and the density can also
become negative in some cases (Kochanek \& Blandford 1987,
Kassiola \& Kovner 1993, Barkana 1998). Because of this, GN 
restricted their model to the small ellipticity of $e=0.3$ measured 
for the lens galaxy light profile by Bernstein et~al.\
(1993). However, the more recent observations by B97 found
that the isophotal ellipticity increases with radius, from 
$0.1$ to $0.4$\,. Furthermore, there is no
guarantee that the dark matter has the same shape as the light
profile, so it is interesting to test the ability of the lensing
data to constrain the dark matter ellipticity directly. 

We use
the SPLS density profile with elliptical isodensity contours,
a model which may be called a softened power-law elliptical mass 
distribution (SPEMD). We calculate the deflection angle and
magnification matrix of this family of models using the fast 
method of Barkana (1998) which avoids the numerical integrations.
We parameterize the SPEMD convergence analogously to the SPLS, as
   \beq \kappa(\vec{\theta}\,)=\frac{\eta}{2} \alpha_E^{2-\eta}
   \left[(x^2/a^2+y^2+\theta_c^2)^{\frac{\eta}{2}-1} \right]\ , \eeq
where we write $\vec{\theta}=(x,y)$, $a$ is the axis ratio (related
to the ellipticity $e=1-a$), and
we assumed a major axis along the $y$-axis. More generally the major 
axis is rotated at an angle $\varphi_a$, which we measure from North
through East, consistent with Bernstein et~al.\ (1993,1997).
The SPEMD thus adds $a$ and $\varphi_a$ to the set of parameters of 
the SPLS. 

We also explore an elliptical density model based on the FGS
profile, keeping the point mass and adding ellipticity parameters
to the King profile. As we did with the SPLS, we first take the 
axisymmetric convergence of the FGS model and substitute $(x^2/a^2+y^2)$
for $r^2$, and then rotate the major axis by an angle $\varphi_a$.
When made elliptical, the convergence (Equation~\ref{kapfgs}) in the
approximation of Young et~al.\ (1981) yields the difference of two
terms, each of which corresponds to the special case of an isothermal 
SPEMD. The deflection angle and magnification of such a softened 
{\it isothermal} elliptical mass distribution has been computed
analytically in terms of complex numbers by Kassiola \& Kovner (1993),
so it is easy to perform lens modeling with the FGS elliptical mass 
distribution, or FGSE.

The lensing galaxy in 0957+561 is a massive galaxy near the center of a
galaxy cluster. Following FGS, we assume that the cluster deflection 
varies on a scale large compared to the image separation, so we expand 
the cluster deflection about the center of the lens galaxy and assume
it has a linear deflection law, $\alpha_i=M_{ij} \theta^j$. The traceless
part of the matrix $M_{ij}$ is a shear $\gamma$ with direction 
$\varphi_{\gamma}$, where
   \beq
   M=\gamma \left(\begin{array}{cc}
   \cos 2\varphi_{\gamma} & -\sin 2\varphi_{\gamma} \\
   -\sin 2\varphi_{\gamma} & -\cos 2\varphi_{\gamma}
   \end{array}\right)\ .
   \eeq
Note that GN denoted the shear angle $\phi$, and we have defined
$\varphi_{\gamma}=-\phi$ for consistency with measuring the position 
angle of a possible corresponding cluster from North through East.
The trace part is a convergence $\kappa$, which corresponds to the 
degeneracy identified by Falco et~al.\ (1985): Given any
lens model, if we multiply the deflection
$\vec{\alpha}(\vec{\theta}\,)$ by the factor
$(1-\kappa)$ and at the same time include a convergence $\kappa$ in
the model, the relative image positions and magnifications remain
unchanged. The time delay changes, however, by the factor $(1-\kappa)$,
inducing an uncertainty in the derived $H_0$ unless
$\kappa$ can be determined. 
GN note that because of this, models really only determine the scaled
shear $\gamma' = \gamma/(1-\kappa)$, and (for a given measured time delay)
a scaled value of h which we denote $h'$, where
$H_0 = 100\, h\, {\rm km\,s^{-1}\,Mpc^{-1}}$
is standard notation and we also have
  \beq
  H_0 = 100\, h' \, (1-\kappa)\, {\rm km\,s^{-1}\,Mpc^{-1}}\ .
  \eeq
In models which include external shear but no explicit 
convergence, the mass of the lens
galaxy is also related to the physical mass by the same factor of
$(1-\kappa)$. This is true for $\alpha_E^{2-\eta}$ of the SPLS and
$\sigma_v^2$ and $M_{\bh}$ of the FGS model.
As noted above, a direct measurement of the 
mass of the lens galaxy or the cluster can determine $\kappa$.
Hereafter we use the symbol $\kappa$ to refer to the convergence
produced by the cluster only.

As an independent attempt to determine $\kappa$, we also model the cluster
as a Singular Isothermal Sphere (SIS) with a variable position, letting
the fit determine the position
as well as the velocity dispersion. For this model, $\rho(r) \propto 1/r^2$,
$\Sigma(\xi) \propto 1/\xi$, and
   \beq
   \vec{\alpha}(\vec{\theta}\,)=b_{\cl} \hat{\theta}'\ , \ \ \ \ \ 
   b_{\cl}=4 \pi\left(\frac{\sigma_{\cl}}{c}\right)^2 \frac{D_L}{D}=
   17\farcs3 \left(\frac{\sigma_{\cl}}{1000\, {\rm km\,s}
   ^{-1}}\right)^2\ ,
   \eeq
where $\sigma_{\cl}$ is the velocity dispersion of the cluster and
$\vec{\theta}'=\vec{\theta}-\vec{\theta}_{\cl}$. The cluster parameters 
in this case are thus $\sigma_{\cl}$ and the coordinates $(x_{\cl},y_{\cl})$ 
of the cluster center $\vec{\theta}_{\cl}$ with respect to the lens galaxy 
position. GN considered this type of profile for the cluster but did not use 
it as part of their lens model. Bernstein et~al.\ (1993) included an
isothermal cluster in some of their models. Some information
can be obtained from lens modeling about the cluster position,
because of the influence of terms of higher order than the shear.
However, fitted models imply a cluster far from the lens 
galaxy and the results are similar to those obtained for a cluster 
at an infinite distance. Therefore in the external
shear model, while $h'$ (which doesn't include the effect of the
cluster convergence) gives only an upper limit to the value of $h$,
we can obtain an estimate of $h$ by assuming an SIS cluster at
infinite distance, i.e. at a distance large compared to the image
separation. In this case, since the external shear model determines
$\gamma'$ while an SIS cluster has $\kappa=\gamma$, we obtain
an estimate for $h$ of 
   \beq h_{\SI}=h' /(1+\gamma')\ . \eeq

We can now count the number of degrees of freedom (ndof) for
various models. We have 8 position constraints (core and jet $(x,y)$ 
positions in images $A$ and $B$, all relative to an observed lens
position) and 6 magnification constraints 
(relative magnification matrix at the brightest jet component
plus the two eigenvalues at the core).
We add an independent core flux ratio constraint, and there is 
one more constraint for non-singular models which produce a third image.
The SPLS and FGS models have 9 parameters : 3 for the lens galaxy 
profile , 2 for the external shear, and 4 for the two source positions.
Two more parameters are added to the elliptical models, and one more
when the SIS cluster is used instead of external shear.
If we interpret the B97 Blob and Knot components
as two additional pairs of lensed images, then
each pair adds 4 position constraints and one flux ratio.
Each pair also adds to the model a source position (2 
parameters) and a variable source redshift, since the redshifts of 
these faint sources have not been measured.
Thus, e.g., the ndof is 6 for the SPLS model fit only to the 
VLBI data and the third image flux limit,
and 8 for the FGSE model fit to the VLBI data, the core flux
ratio, and the HST Knots and Blobs of B97.

\section{Results \& Discussion}
\label{09s5}

In this section we apply the lens models defined in \S 4
to the constraints described in \S\S 2 and 3, and discuss
the results which are summarized in Table~4.
For each model, we use 
$\bar{\chi}^2$ to denote the reduced $\chi^2$,
and estimate $95\%$ confidence bounds as in GN, 
from the conservative condition $\Delta\chi^2=4\bar{\chi}^2$.
Confidence ranges are included for the FGSE model and all $H_0$ values,
to illustrate the scale of our uncertainties.
We also assume an Einstein-deSitter
$\Omega=1$ cosmology in deriving $H_0$ values.
The effects of this assumption are small for standard cosmologies,
e.g., an open $\Omega=0.3$ universe increases
the $H_0$ estimate by $\sim6\%$, 
while a flat $\Omega_{\rm matter}=0.3$ universe with a cosmological constant
yields an increase of only $\sim4\%$. 
Finally, we use the Kundi\'c et~al.\ (1997) time delay measurement
of 417$\pm$3~days, throughout.

We begin with the axisymmetric models for the lens galaxy together
with the external shear model for the cluster, and fit to the 
full set of constraints (Table~3). 
The first two columns of Table~4 show the best-fit parameters
for the SPLS and FGS models (for $\kappa=0$). 
Note that some of the parameter values using our
corrected constraints differ substantially
from the corresponding results of GN. 
However, the new constraints are very poorly fit,
with $\bar\chi^2$ values over three times those of GN.

The lens galaxy is observed to be elliptical (B97),
and when we add ellipticity as a parameter
the lens models gain great flexibility. 
The $\bar\chi^2$ values are considerably lower for the 
elliptical SPEMD and FGSE models (see Table~4), 
and are comparable to the GN goodness-of-fit estimates. 
As a check, we tried using the SPLS profile
with elliptical isopotentials as used by GN, instead of the
elliptical isodensity contours of the SPEMD, and this fit gave
similar parameter values to the SPEMD but with a larger 
$\bar\chi^2$ of 14. 
The $H_0$ estimates are very different for the FGSE and SPEMD models,
but the FGSE has a much lower $\bar\chi^2$. 
The FGSE model also provides a closer match to the observed galaxy orientation
($\varphi_{\rm obs}\approx40^{\circ}$, 
with a scatter of $\sim10^{\circ}$, see B97).
Mass and light tend to align to within $\sim10^{\circ}$ in other
lens systems, once external tides are accounted for 
(Keeton, Kochanek, \& Falco 1998).
Unfortunately, the two sources of asymmetry in both models 
are nearly degenerate, which tends to increase the error ranges on
all parameter estimates. FGSE models with zero external shear
are within our 2$\sigma$ range, implying that there is
no lower limit on $\gamma'$, and that 
$\phi_{\gamma}$ is undetermined.
For both the FGSE and SPEMD models, the ellipticity is high,
and for the FGSE model $h_{\SI}$ increases steadily as $a$ decreases.
If we constrain the ellipticity to equal the highest value 
observed for the light (i.e., axis ratio set to 0.6, see B97), 
we decrease $h_{\SI}$ to 1.01 for the FGSE model, yielding
$\bar{\chi}^2=8.0$. The SPEMD with $a=0.6$ yields 
$h_{\SI}=0.626$ with $\bar{\chi}^2$=10.4\,. 

A more accurate description of the cluster contribution could remove much of the
uncertainty in the models.  Note that the FGSE and SPEMD models make very
different $H_0$ predictions, with the SPEMD requiring four times as much 
external shear. 
We can model the cluster contribution
by replacing the external shear with a simplified cluster mass model
(see,~e.g., Kochanek 1993). 
For example, our FGSE+CL model combines a shear-less FGSE model with a movable
SIS cluster mass distribution. 
The overall $\bar{\chi^2}$ is comparable to that of the FGSE but 
the estimated values for some of the parameters are substantially changed.
The $h$ estimate is very similar to that for the FGSE, 
but the uncertainty has increased from $\sim20\%$ to $\sim30\%$,
demonstrating the sensitivity of the result to the assumed model.
Figure~2 shows the appearance of the source and image planes for this model.
We compare in Figure~3
the estimates from the FGSE and FGSE+CL lens models with the estimates
from observations of the cluster center and velocity dispersion.
Both of the observed cluster positions (Fischer et~al.\ 1997)
are offset from $G1$ in the same direction, 
and agree approximately with the positions 
from the two lens models. Note that the external shear model does
not depend on whether the cluster lies to the East or West of $G1$,
but the SIS cluster model breaks this degeneracy in favor of the 
observed direction.  Still, the observational uncertainties 
encompass most of the models within our 2$\sigma$ contour.
Likewise, the observed velocity dispersions 
(Fischer et~al.\ 1997; Garrett et~al.\ 1992; Angonin-Willaime et~al.\ 1994)
agree with the model estimates, and are too imprecise to distinguish
between the models.
The error estimates on cluster parameters also depend strongly
on the effective $\gamma'$, which produces large uncertainties for FGS
models due to the weaker cluster contribution.
Nevertheless, it is clear that more precise measurements of the cluster
properties should provide significant constraints on the lens models.
For example, if we assume a cluster velocity dispersion of 
715~km/s and place the cluster center at the same distance
($32\farcs2$)
as the ``galaxies'' position, derived from number counts (see Figure~3),
we obtain $h=1.05$ with $\bar{\chi}^2=49.3/9=5.5$.
The same velocity dispersion and a distance of $22\farcs2$,
corresponding to the ``weak'' position from weak lensing,
lowers $h$ further to 0.86 with $\bar{\chi}^2=82.3/9=9.1$.
More precise cluster measurements may also result in $G1$ mass estimates
which are more typical of a massive elliptical galaxy. 
For the clusters at the ``galaxies'' and ``weak'' positions,
the corresponding $G1$ velocity dispersions $\sigma_v$ are
378~km/s and 331~km/s, respectively. 

Thus far, we have assumed an SIS cluster profile, 
but other profiles would yield different estimates of $h$.
We can explore the effect of different cluster profiles
by approximating the cluster as an external
shear $\gamma$ and convergence $\kappa$. Then, given 
$\gamma'$ and $h'$, 
there is a relation between $\gamma$ and $\kappa$ 
for each cluster profile and position 
which allows us to estimate both, and thus $h=h' (1-\kappa)$.
Figure~4 illustrates the effect on the estimate of $h$
of using different profiles
for the simple, hypothetical case 
of $\gamma'=0.20$ and $h'=1.00$\,.
If we assume that the cluster is spherically symmetric
and described by the SPLS profile, the estimate
of $h$ depends on the cluster power-law index $\eta$ and on
the distance to the cluster from the lens galaxy in
units of the cluster core radius. 
If the cluster is singular, even large deviations in
$\eta$ from the isothermal value of unity have a small effect
on the estimated $h$. 
This insensitivity results from the estimates of $\kappa$ being
small, so large
fractional changes in $\kappa$ produce smaller fractional
changes in $(1-\kappa)$. On the other hand, if $G1$
is within a few cluster core radii of the cluster center, 
then the SPLS profile approaches a constant density sheet 
(corresponding to a $1/r$ density profile in 3 dimensions);
so with $\gamma'$ fixed, $\kappa$ is driven toward 1 and $h$
decreases. The observations of Fischer et~al.\ (1997)
imply a cluster core radius of $5\pm5\arcsec$,
for an isothermal cluster.
This accuracy is insufficient for use in lens modeling;
the determination of the cluster's mass distribution
from weak lensing is difficult
because of the insufficient number of faint background
sources in the small central area of the cluster.
However, a more precise determination 
of the cluster's center and mass profile
may allow us to distinguish between some models, 
and thus further reduce the model uncertainties.

As noted in \S \ref{09s3}, 
the center of $G1$ has been estimated from 
VLBI (Gorenstein et~al.\ 1983)
and HST (B97) observations;
we denote these positions as $G'$ and $G1$, respectively. 
If we substitute the $G'$ position with its 1~mas standard errors
for the $G1$ position in Table~3,
then the parameter estimates are
almost unchanged for the SPEMD and FGSE models.
Of course, it is not clear how close the center of light 
--- radio or optical --- is to the mass center of the galaxy.
It is correspondingly unclear what standard deviation should be used
for these possible separations.
Moreover, the mass distribution
may not follow the shape of the light distribution,
which has an ellipticity that varies with radius (B97).
As an extreme case, we can assume an effectively infinite 
position uncertainty, by removing the lens position constraint.
The resultant fit to an SPEMD model has 
$\bar{\chi}^2=57.9/7=8.3$, with $a=0.48$, 
$\varphi_a=-16\deg$, $h_{\SI}=0.796$, and 
the estimated lens center is displaced by $(21,-78)$~mas 
from $G1$'s optical center.
The corresponding values for the FGSE model are
$\bar{\chi}^2=22.9/6=3.8$, 
with $a=0.53$, $\varphi_a=-1\deg$, $h_{\SI}=1.01$ 
and the lens displaced by $(45,-44)$~mas. 
These large estimated lens displacements from the center of light
call for a general study of how much the mass and light centers
should differ in the cores of elliptical galaxies.
If we allow the center of mass for $G1$ to
be as far as $\sim80$~mas away from the optical center, 
then the estimated $h_{\SI}$ changes by $\sim20\%$.
Another indication of uncertainty in our models 
is that image $B$ is well inside the effective radius of $G1$
($R_{\rm eff}\sim4\farcs5$, Bernstein et~al.\ 1993),
while image $A$ lies just outside this radius.
Although we might expect that the galaxy mass is dominated
by stars out to the distance of $B$, there may be 
a significant dark matter contribution at the radius of $A$,
requiring a more complicated model.

Since our full set of constraints supplies a fairly large 
number of degrees of freedom, we can explore the robustness of
the results by observing the effect of removing individual
constraints. If, e.g., we use the FGSE model without including the HST Blobs
(but including the Knots),
we find $\bar{\chi}^2=9.1/6=1.5$ with $a=0.15$, $\varphi_a=73\deg$,
and $h_{\SI}=0.340$.  This estimated ellipticity 
is much higher than that of the light distribution,
which suggests that the
FGSE model is not well constrained without the Blobs.
By contrast, the Knots are only weak constraints due
to the large errors associated with their positions and fluxes.
The results are thus clearly sensitive
to which constraints are included.

As another test of robustness, we used the FGSE model with the full set
of constraints but we recomputed the VLBI constraints requiring
the $B$~image component shapes also to agree 
with those of the $A$~image through the 
spatially-varying magnification transformation.
The resulting parameter values and uncertainties
are almost identical to the FGSE results in Table~4,
but with $\bar{\chi}^2=10.0$, higher than before.

Our models also yield estimates of the distances to the HST objects.
For example, the FGSE $f_{\rm blob}$ and $f_{\rm knot}$ values
correspond to redshifts of $z_{\rm blob}=1.64$ 
and $z_{\rm knot}=3.54$ for $\Omega=1$.
Assuming an open $\Omega=0.3$ universe 
or a flat universe with $\Omega_{\rm matter}=0.3$
increases $z_{\rm knot}$ by about 50\% and 20\% respectively,
but for both cases $z_{\rm blob}$ increases by only a few percent.
The allowed 2$\sigma$ ranges are wide, 
(e.g., $1.15 < z_{\rm blob} < 2.42$ and $1.82 < z_{\rm knot} < 12.2$
for $\Omega=1$)
so the Blob and Knot sources could be at the same distance
for any of these cosmologies.
Note that all of our models predict additional fainter counterimages
of the Knot source close to the observed Blob images (Figure~2).
Such counterimages may have been marginally detected (Avruch et~al.\ 1997) 
in the HST images.  If the Knot and Blob sources are 
physically associated, the additional Knot images could provide
more stringent lensing constraints.

Large-scale mass fluctuations along the line-of-sight to 
Q0957+561 can produce an additional source
of uncertainty in the $H_0$, which will be important 
if the cluster is properly modeled.
Barkana (1996) shows that large-scale structure affects the determination of
$H_0$ with an uncertainty 
$\Delta_1$, but for models which are normalized to the $G1$
velocity dispersion, the velocity dispersion effectively 
constrains part of the effect of large-scale structure, and
a smaller uncertainty $\Delta_2$ is left over. Given the source 
and lens redshifts for Q0957+561, a suite of models 
for the power spectrum of large-scale structure
(Barkana 1996 and Figure~2 of Keeton et~al.\ 1997)
yields typical $2\sigma$ uncertainties of
$\Delta_1=9.8\%$ and $\Delta_2=5.5\%$.

It is often argued that $H_0$ estimates from the Q0957+561 time delay
are less reliable than those obtained from other lensed systems,
because the cluster contribution is important.
But most of the other systems for which a time delay has been measured
have significant lensing contributions from a nearby group of galaxies.
Although the cluster in Q0957+561 is more dominant 
than a smaller group would be,
its mass distribution can, in principle, be directly measured.
Only limited information on the cluster is presently available,
but future prospects are promising for deeper weak
lensing observations and for high-resolution X-ray measurements
with the AXAF satellite. 
It may even be possible to check the cluster for 
dominant substructure with AXAF. 
Another possibility for determining indirectly the cluster contribution
to lensing was suggested by Romanowsky \& Kochanek (1998),
who used the velocity dispersion measurement of $G1$
to estimate its mass distribution via detailed modeling 
of the stellar velocity distribution. 
Unfortunately, the result depends on the
mass distribution near the center of $G1$, which is poorly
determined by lens models, especially for the FGS profile
which has a large point mass at the center. A possible
solution is to construct a lens model which follows the
light shape near the center but becomes an independent dark
matter halo farther out. The present
data cannot constrain the additional parameters necessary
for such a lens model.

\section{Conclusions}

We have used improved data in the analysis of the gravitational lens system
Q0957+561.  We re-analyzed the VLBI data of G94
with corrected numerical procedures, and obtained new estimates
for the components and spatial gradients
of the relative magnification matrix between the $A$ and $B$ images.
We also included new lensing constraints from recently discovered
optical components (B97).
The VLBI and optical constraints were used to determine 
more elaborate lens models than had been previously explored.
In particular, we considered models with two sources of asymmetry:
ellipticity in the lens galaxy and external shear 
from the surrounding cluster of galaxies.

Models with an axially symmetric lens are unable to fit the data,
yielding $\bar{\chi}^2=23$ for the SPLS and $\bar{\chi}^2=27$
for the FGS model (all with the optical $G1$ lens position). Adding 
ellipticity as a model parameter leads to $\bar{\chi}^2=9.9$ 
for the SPEMD model and $\bar{\chi}^2=6.0$ for the FGSE model. The $H_0$
estimates derived from these two models differ substantially, with 
$h_{\SI}=0.61^{+.18}_{-.16}$ 
and $h_{\SI}=1.23^{+.22}_{-.23}$, respectively, 
where the uncertainties correspond to two standard deviations
when an SIS cluster is used to represent the external shear.
The two models can be distinguished,
in principle, since they differ greatly 
in predicting the lens ellipticity direction and the magnitude of the 
cluster shear.  Direct measurements of the cluster
mass distribution thus have great potential.

The simple lens models that we have considered cannot be
uniquely constrained by VLBI measurements alone.
The HST Blobs and Knots (B97) have lines of sight
that are far away from those for the $A$ and $B$ images,
and could eliminate highly elliptical lens models that are permitted
by the basic VLBI and core flux constraints. 
The discovery of more background sources in the field, 
or other extended radio structures (see,~e.g., Avruch et~al.\ 1997),
might eventually distinguish between models, 
and thus narrow the allowed range of $H_0$.
New structures may also provide enough constraints to permit
the application of more complicated and realistic mass models 
which can account for all of the observations. A reliable 
measurement of $H_0$ may be achievable by combining the results 
from several such well-studied lensed systems.


\acknowledgements
We are truly grateful to Mike Garrett for sending us the
VLBI data and to Gary Bernstein for making some of the
HST results available to us before publication. We thank
Paul Schechter and Chris Kochanek for many valuable discussions
and Ed Bertschinger, Peter Schneider, Simon White, and Avi Loeb
for helpful discussions. RB acknowledges support by
Institute Funds and by NASA grant NAG5-2816. JL is grateful for
support from NSF grant AST93-03527 and from the NASA/HST grant GO-7495.
This research was supported by the Smithsonian Institution.

\section{APPENDIX}
\label{09ap}

The two images $A$~and $B$, and hence their VLBI components,
are related by a relative magnification mapping.
Up to an irrelevant translation, we can describe the mapping 
from image $A$ to image $B$ by:
   \beq \bf{x}^B_{i}=\bf{M}^{BA}_{ij}\bf{x}^A_{j}
   +\frac{1}{2}\bf{\partial M}^{BA}_{ijk}\bf{x}^A_{j}\bf{x}^A_{k}\ 
   \label{magmap}, \eeq
where repeated indices are summed. Here, 
$\bf{x}^A$ and $\bf{x}^B$ are the respective positions
in the $A$ and $B$ images, referred to the origins at 
the centers of their respective cores;
$\bf{M}^{BA}$ 
is the $2\times 2$ relative magnification matrix, evaluated at the 
center of $A_1$;
and $\bf{\partial M}^{BA}$ is the tensor that represents
the next order term in a Taylor series expansion
(this tensor, by definition, 
is symmetric with respect to its last two indices).
There are thus 4 independent parameters that define 
$\bf{M}^{BA}$ and 6 that define $\bf{\partial M}^{BA}$. The data,
however, which are confined to a relatively small region, are
sensitive only to the magnification derivatives along the jet direction,
and weakly even to these.  When we attempt estimates which include
all 6 independent components of $\bf{\partial M}^{BA}$ the
resultant values for most components are much higher than is 
physically reasonable. 
Thus, we remove the parameters to which our data are insensitive.

The mapping of Equation~\ref{magmap} describes 
the expansion of the relative magnification matrix 
$\bf{M}^{BA}(\bf{x}^A)$ about the origin:
   \beq \bf{M}^{BA}_{ij}
   (\bf{x}^A)\equiv\frac{\partial \bf{x}^B_i}{\partial \bf{x}^A_j}
   =\bf{M}^{BA}_{ij}+\bf{\partial M}^{BA}_{ijk}\bf{x}^A_{k}\ .
   \label{map2} \eeq
For such a position-dependent magnification matrix, 
a Gaussian representation of flux density components in the $A$ jet
is no longer mapped to a corresponding Gaussian representation
in the $B$ jet.  In our analysis, however, we ignore the 
variation of $\bf{M}^{BA}$ over the extent of a component;
thus we map component $A_4$, for example,
to $B_4$ using a relative magnification matrix,
from Equation~\ref{map2}, evaluated at the center of $A_4$.
This approximation provides another reason for our
treating the error estimates conservatively. Following
Gorenstein et~al.\ (1988) and G94, we decompose the matrix
$\bf{M}^{BA}$ into its eigenvalues ($M_1$ and $M_2$) and
the corresponding position angles of the eigenvectors
($\phi_1$ and $\phi_2$). The matrix can be represented as
   \beq {\bf M^{BA}}=M_1 {\bf E}(\phi_1,\phi_2)+ M_2 {\bf E}(\phi_2,
   \phi_1)\ , \eeq
where 
   \beq \bf{E}(\rho,\sigma)=
   \left[\begin{array}{c}
   \cos \rho \\ \sin \rho
   \end{array}\right] \cdot
   \left[\begin{array}{cc}
   -\sin \sigma & \cos \sigma
   \end{array}\right] \cdot \csc(\rho-\sigma)\ , \eeq
in the notation of G94, but without rotating
coordinates to align with the $A$ jet as do G94. 
Because of the limited sensitivity of our data, 
we restrict our estimation to a 
subset of the $\bf{\partial M}^{BA}$ parameters.
We fix $\phi_1$ and 
$\phi_2$ to be constant along both the jet axis and 
the perpendicular direction, which is slightly 
different from the procedure of G94. 
Thus, we fix four $\bf{\partial M}^{BA}$ components through:
   \begin{eqnarray}
   \bf{\partial M}^{BA}_{221}&=&\frac{\bf{M}^{BA}_{21}}
   {\bf{M}^{BA}_{12}}\bf{\partial M}^{BA}_{122}\ , \nonumber \\
   \bf{\partial M}^{BA}_{211}&=&\frac{\bf{M}^{BA}_{21}}
   {\bf{M}^{BA}_{12}}\bf{\partial M}^{BA}_{121}\ , \nonumber \\
   \bf{\partial M}^{BA}_{111}&=&\frac{\bf{M}^{BA}_{21}}
   {\bf{M}^{BA}_{12}}\bf{\partial M}^{BA}_{122}-
   \frac{\bf{M}^{BA}_{22}-\bf{M}^{BA}_{11}} {\bf{M}^{BA}_{12}}
   \bf{\partial M}^{BA}_{121}\ , \label{restrict} \\
   \bf{\partial M}^{BA}_{222}&=&\bf{\partial M}^{BA}_{121}+
   \frac{\bf{M}^{BA}_{22}-\bf{M}^{BA}_{11}}
   {\bf{M}^{BA}_{12}}\bf{\partial M}^{BA}_{122}\ , \nonumber 
   \end{eqnarray}
leaving only two independent components of the derivative matrix.

\vfill\eject

\begin{figure}
\epsscale{0.8}
\plotone{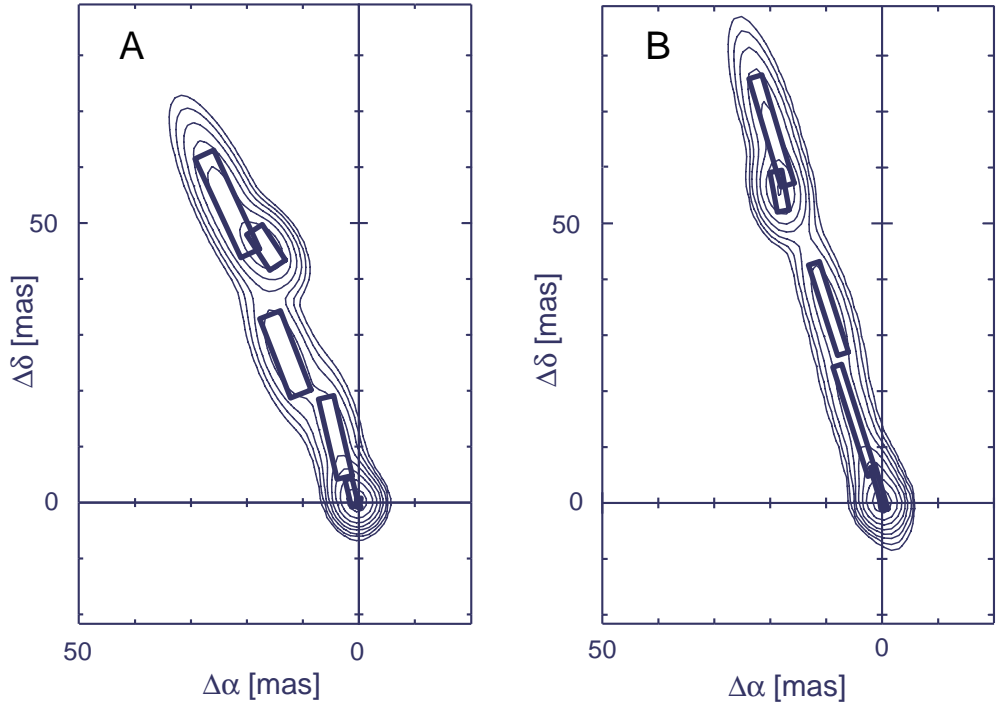}
\caption{The surface brightness distributions for the Gaussian
         VLBI components (see Table~1), convolved with a
         circular 4-mas beam, with image $A$ on 
         the left and $B$ on the right. 
         Position coordinates, $\Delta\alpha$ and $\Delta\delta$,
         are in milliarcsec East and North on the sky,
         relative to the respective core components.
         Rectangles represent
         the axes and position angles of individual flux components,
         $A_i$ and $B_i$, where $i$ increases with 
         distance from the core along the jet.
         The contours increase by factors of two from
         0.55\,mJy and 0.39\,mJy per beam, 
         for the A and B image maps respectively.
         }
\end{figure}

\begin{figure}
\epsscale{0.7}
\plotone{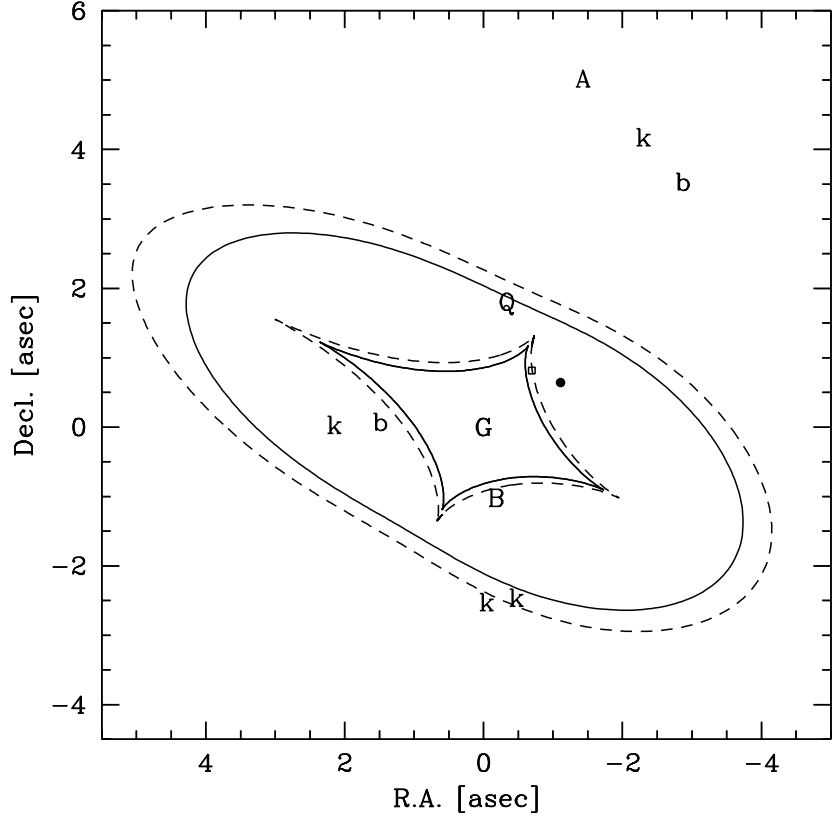}
\caption{Source and image planes for the FGSE+CL model. 
    We show the source positions for the Blobs ($\bullet$), the Knots
    ($\Box$), and the quasar ($Q$), as well as corresponding image
    positions of the Blobs ($b$), the Knots ($k$), and the quasar
    ($A$ and $B$), all relative to the lens galaxy ($G$).
    The caustics (inner pair) and critical curves (outer pair) are shown
    for the $z=1.4$ quasar (solid lines) 
    and the higher redshift Knot source (dashed lines).
    The corresponding curves for the Blob source would fall 
    between them. The Knot source is just inside its caustic, 
    producing the close pair of observed Knot images,
    as well as two faint counterimages near the two Blobs.
    }
\end{figure}

\begin{figure}
\epsscale{0.7}
\plotone{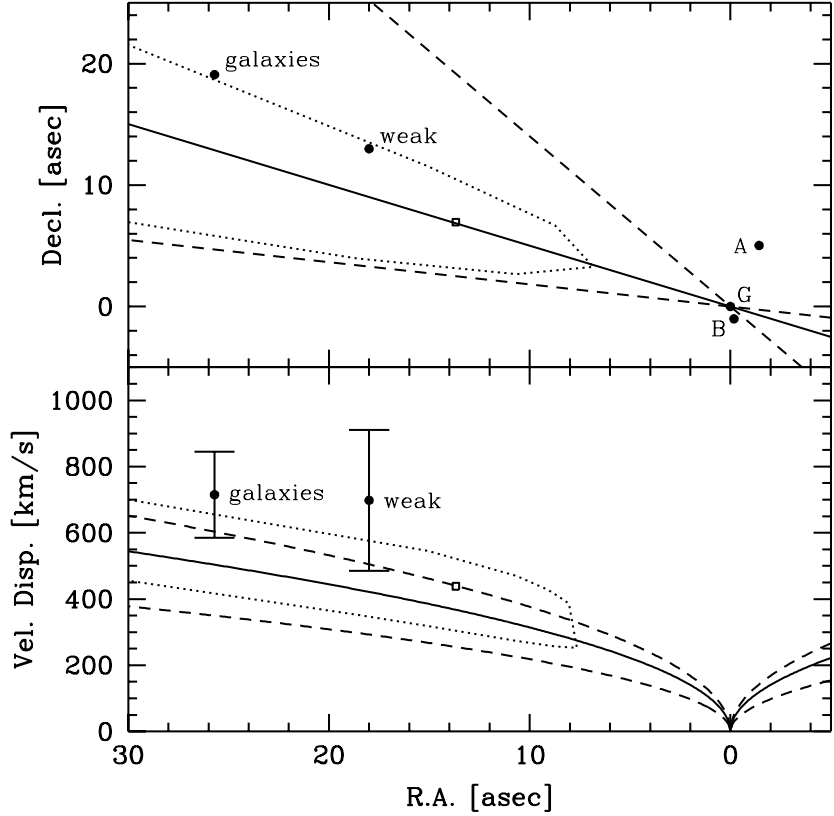}
\caption{Cluster properties for the FGSE and FGSE+CL models compared
 to observations.
     The top panel shows the cluster position relative to the lens
 galaxy $G1$ (labeled $G$ in the Figure).
     The quasar image locations ($A$ and $B$) are also shown.
     From the FGSE model with shear we derive the direction to
     the cluster center (solid line) with a $2\sigma$ range (dashed
 lines), assuming only a spherical cluster.	
     From the FGSE+CL model, we derive the
     cluster center ($\Box$) and its $2\sigma$ contour (dotted curve).
     Also shown are the estimated ``weak'' cluster center from weak lensing,
     and the ``galaxies'' position from galaxy number counts 
     (Fischer et~al.\ 1997). These observed positions have large uncertainties
 which are not shown, roughly $28\arcsec$ and $21 \arcsec$
respectively, at $1\sigma$.
     The bottom panel compares the predicted cluster velocity dispersion
     to the Fischer et~al.\ estimate from weak lensing (``weak''),
     and the measured dispersion (``galaxies'') for 21 probable
 cluster members (Garrett et~al.\ 1992; Angonin-Willaime et~al.\ 1994).
 Observed errors are again $1\sigma$ ranges.
     The effective velocity dispersion of the FGSE plus shear model
     is shown as a solid curve, with dashed 2$\sigma$ error range curves,
     assuming the SIS relation $\gamma=\kappa$,
     and the FGSE+CL best-fit and 2$\sigma$ contour are shown as above.
     Unlike Table 4, in this figure all $2\sigma$ errors from lens
     models are the formal errors of
     $\Delta \chi^2=4$ for one variable and $\Delta \chi^2=6.17$ for
     two variables.}
\end{figure}

\begin{figure}
\epsscale{0.7}
\plotone{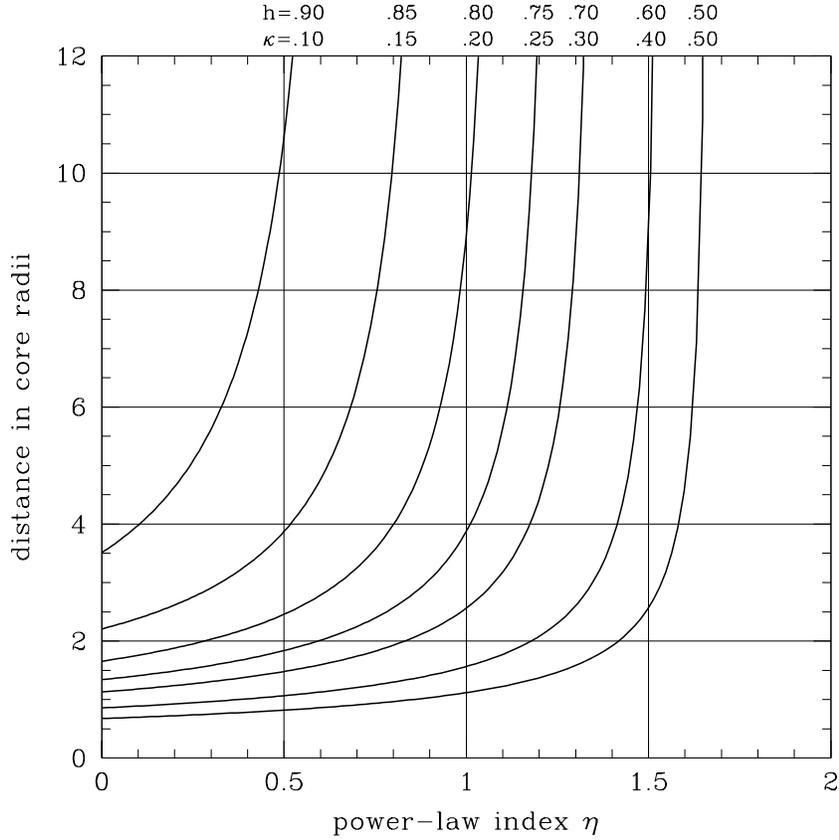}
\caption{Contours of constant convergence and corresponding
    values of $h$, shown as a function of cluster properties. The 
    cluster is assumed to have an SPLS profile, with its $\eta$ as
    the $x$-axis. The $y$-axis is the distance of the cluster from the 
    lens galaxy divided by the cluster core radius. Values of $\kappa$
    are indicated at the top for each contour, with the corresponding
    value of $h$ for the hypothetical case of $\gamma'
    =0.20$ and $h'=1.00$. }
\end{figure}

\vfill\eject

\begin{deluxetable}{lcccccc}
\tablecaption{VLBI Component Parameters}
\tablenum{1}
\tablehead{
\colhead{Component} & \colhead{Flux} & \colhead{$\Delta \alpha$}   
& \colhead{$\Delta \delta$} & \colhead{Maj. Axis} &
\colhead{Axis Ratio} & \colhead{Pos. Ang.} \\
\colhead{} & \colhead{(mJy)} &
\colhead{(mas)} & \colhead{(mas)} &
\colhead{(mas)} & \colhead{} &
\colhead{$(\deg)$} }
\startdata
%
%
$A_1$ & $8.7\pm1.0$ & 0 & 0 & $1.6\pm0.2$ & $0.51\pm0.15$ & 
$-1\pm 8$ \nl 
$A_2$ & $7.6\pm1.1$ & $1.3\pm0.1$ & $1.8\pm0.3$ & $5.0\pm0.4$ & 
$0.26\pm0.05$ & $13\pm3$ \nl
$A_3$ & $3.9\pm0.5$ & $4.2\pm0.2$ & $11.7\pm0.8$ & $14.7\pm2.1$ & 
$0.19\pm0.03$ & $13\pm2$\nl
$A_4$ & $4.7\pm0.3$ & $13.1\pm0.2$ & $26.5\pm0.4$ & $15.2\pm1.2$ & 
$0.26\pm0.03$ & $21\pm2$\nl
$A_5$ & $9.1\pm0.4$ & $16.6\pm0.1$ & $45.6\pm0.1$ & $7.5\pm0.3$ & 
$0.44\pm0.02$ & $33\pm1$\nl
$A_6$ & $8.8\pm0.5$ & $23.5\pm0.3$ & $53.4\pm0.5$ & $19.5\pm1.0$ & 
$0.19\pm0.01$ & $25\pm1$\nl
\tableline
$B_1$ & $7.2\pm0.5$ & 0 & 0 & $2.0\pm0.2$ & $0.37\pm0.10$ & $9\pm3$ 
\nl 
$B_2$ & $5.5\pm0.6$ & $0.5\pm0.2$ & $2.6\pm0.5$ & $8.0\pm0.8$ & 
$0.12\pm0.05$ & $16\pm1$ \nl
$B_3$ & $2.8\pm0.4$ & $4.9\pm0.4$ & $14.8\pm1.0$ & $20.5\pm3.7$ & 
$0.08\pm0.03$ & $18\pm2$\nl
$B_4$ & $2.8\pm0.2$ & $9.6\pm0.2$ & $34.8\pm0.7$ & $17.0\pm1.8$ & 
$0.12\pm0.03$ & $18\pm2$\nl
$B_5$ & $5.8\pm0.3$ & $18.3\pm0.1$ & $55.8\pm0.2$ & $7.2\pm0.4$ & 
$0.30\pm0.02$ & $11\pm1$\nl
$B_6$ & $4.8\pm0.4$ & $19.8\pm0.3$ & $66.5\pm0.9$ & $20.2\pm1.5$ & 
$0.12\pm0.01$ & $17\pm1$\nl
\enddata
\tablecomments{$B$~image fluxes and positions are not independently
   determined, they are directly related to the
   corresponding $A$~image values through the magnification transformation
   given in Table~2.
   We give the estimated $1\sigma$ errors on the parameters. 
   Positions are measured relative to the respective core components. 
   Position angles are measured in the direction North through East.\\
   }
\end{deluxetable}
 
\begin{deluxetable}{lrrrrrrr}
\tablenum{2}
\tablecaption{VLBI Constraints}
\tablehead{
\colhead{Quantity} & \colhead{Measured}&
\multicolumn{6}{c}{correlation coefficients} \\ \tableline \\
\colhead{} & \colhead{} & \colhead{$A_5\Delta \alpha$} 
& \colhead{$A_5\Delta \delta$} & \colhead{$B_5\Delta \alpha$} & 
\colhead{$B_5\Delta \delta$} & \colhead{} & \colhead{}}
\startdata
%
%
$A_5\Delta \alpha$ (mas) & $16.6\pm 0.1$ & 1.00 & & & & & \nl
$A_5\Delta \delta$ (mas) & $45.6\pm 0.1$ & 0.26 & 1.00 & & & & \nl
$B_5\Delta \alpha$ (mas) & $18.32\pm 0.07$ & --0.40 & 0.27 & 1.00 & & & \nl
$B_5\Delta \delta$ (mas) & $55.8\pm 0.2$ & --0.12 & 0.19 & 0.36 & 1.00 & & \nl
\tableline
\colhead{} & \colhead{} & \colhead{$M_1$ at $A_5$} & \colhead{$M_2$ at
$A_5$} & \colhead{$\phi_1$} & \colhead{$\phi_2$} & \colhead{$M_1$ at 
$A_1$} & \colhead{$M_2$ at $A_1$} \nl
\tableline
$M_1$ at $A_5$ & $1.15\pm 0.03$ & 1.00 & & & & & \nl
$M_2$ at $A_5$ & $-0.56\pm 0.03$ & 0.05 & 1.00 & & & & \nl
$\phi_1$ $(\deg)$ & $18.76\pm 0.04$ & --0.27 & 0.25 & 1.00 & & & \nl
$\phi_2$ $(\deg)$ & $107\pm7$ & --0.34 & --0.74 & --0.16 & 1.00 & & \nl
$M_1$ at $A_1$ & $1.27\pm0.03$ & --0.95 & --0.02 & 0.20 & 0.16 & 1.00 & \nl
$M_2$ at $A_1$ & $-0.58\pm0.04$ & --0.06 & 0.985 & 0.31 & --0.70 & 0.09 & 1.00 \nl
\enddata
\tablecomments{Results of simultaneous fitting to the $A$
and $B$ images, where the positions and fluxes of the $B$ 
jet components are fixed by the $A$ jet components and the $A\rightarrow B$ 
mapping, but the shapes are not. We give the estimated $1\sigma$ errors 
on the parameters, the normalized position correlation coefficients, 
and the magnification correlation coefficients. 
The component positions for $A_5$ and $B_5$, from Table~1, 
are included as separate constraints as they provide the most
precise information on the jet structure.
}
\end{deluxetable}

\begin{deluxetable}{lccl}
\tablenum{3}
\tablecaption{``Full set'' of constraints}
\tablehead{ \colhead{Constraint} & \colhead{Value(s)} &
\colhead{${\rm N_{const}}$}
& \colhead{Comments} }
\startdata
Magnification matrix & See Table 2 & 6 & from fit to VLBI data \nl
$A_5$ position & See Table 2 & 2 & from fit to VLBI data \nl
$B_5$ position & See Table 2 & 2 & from fit to VLBI data \nl
$A_1$ core position & $(-1\farcs25254,6\farcs04662)\pm 0\farcs00004$
& 2 & corrected from FGS \nl
$B_1/A_1$ magnif & $0.747 \pm 0.015$ & 1 & Conner et~al.\ 1992 \nl
$C/B_1$ magnif & $< 1/30$ & $1^{*}$ & as in GN; $^{*}$non-FGS only \nl
lens center & $(0\farcs 1776, 1\farcs 0186)\pm 0\farcs0035$ &
2 & G1, from HST, B97 \nl
Blob East pos & $(1\farcs 72,0\farcs 98)\pm 0\farcs 05$ & 2
& Blob 2, B97 \nl
Blob West pos & $(-2\farcs 70,4\farcs 48)\pm 0\farcs 05$ &
2 & Blob 3, B97 \nl
Blobs magnif ratio & $-1.3\pm 0.14$ mag & 1 & Blobs 2/3 flux, B97 \nl
Knot East pos & $(0\farcs13,-1\farcs 53)\pm 0\farcs 05$ & 2
& Knot 1, B97 \nl
Knot West pos & $(-0\farcs29,-1\farcs 41)\pm 0\farcs 05$ & 
2 & Knot 2, B97 \nl
Knot magnif ratio & $-0.3\pm 0.7$ mag & 1 & Knots 1/2 flux, B97
\enddata
\tablecomments{All positions are offsets ($\Delta\alpha$,$\Delta\delta$) from the $B_1$ core position.}
\end{deluxetable}

\begin{deluxetable}{lccccc}
\tablenum{4}
\tablecaption{Best fit model results}
\tablehead{ \colhead{Parameter} & \colhead{SPLS} & \colhead{FGS}
& \colhead{SPEMD} & \colhead{FGSE} & \colhead{FGSE+CL} }
\startdata
\sidehead{Lensing galaxy:}
$x_{\rm lens}-x_{G1}$ & 40.4\,mas & 16.4\,mas & 6.3\,mas & $4.4^{+17}_{-16}$\,mas & 2.5\,mas \nl
$y_{\rm lens}-y_{G1}$ & 0.0\,mas & 2.5\,mas & --1.7\,mas & $-0.2^{+17}_{-17}$\,mas & --0.2\,mas \nl
$\theta_c$ & $0\farcs00$ & $2\farcs58$ & $0\farcs17$ & 
$1\farcs59^{+1.6}_{-0.2}$ & $1\farcs39$ \nl
$\alpha_E$ or $\sigma_v$ & $2\farcs45$ & 378\,km/s & $3\farcs98$ & 
$447^{+350}_{-54}$\,km/s & 404\,km/s \nl
$\eta$ or $M_{\bh}$ & 1.16 & 107\,$\times10^9$M$_{\odot}$ & 1.01 &
92.5$^{+14}_{-26}$\,$\times10^9$M$_{\odot}$ & 69.9\,$\times10^9$M$_{\odot}$ \nl
$a$        & \nodata & \nodata & 0.469 & $0.445^{+.15}_{-.32}$ & 0.505 \nl
$\varphi_a$& \nodata & \nodata & $123\fdg8$ & $67\fdg3^{+18}_{-19}$ 
& $61\fdg 9$ \nl
\sidehead{Cluster contribution:}
$\gamma'$ & 0.258 & 0.209 & 0.364 & $0.083^{+.10}_{-\ldash}$ & \nodata \nl
$\varphi_{\gamma}$ & $66\fdg 4$ & $65\fdg 0$ & $51\fdg 9$ & $63\fdg 4
\pm \ldash$ & \nodata \nl
$\sigma_{\cl}$ & \nodata & \nodata & \nodata & \nodata & 439\,km/s \nl
$x_{\cl}-x_{G1}$ & \nodata & \nodata & \nodata & \nodata & $13\farcs7$ \nl
$y_{\cl}-y_{G1}$ & \nodata & \nodata & \nodata & \nodata & $6\farcs9$ \nl
\sidehead{HST sources:}
$f_{\rm blob}$ & 1.06 & 1.02 & 1.03 & $1.05^{+.10}_{-.13}$ & 1.05 \nl
$f_{\rm knot}$ & 1.08 & 0.95 & 1.09 & $1.22^{+.12}_{-.14}$ & 1.12 \nl
\sidehead{General:}
ndof & 11 & 10 & 9 & 8 & 7 \nl
$\bar{\chi}^2$ & 23 & 27 & 9.9 & 6.0 & 5.9 \nl
$h$ or $(h_{\SI})$ & $(0.619^{+.06}_{-.03})$ & $(0.657^{+.11}_{-.28}$)
& $(0.629^{+.16}_{-.18})$ & $(1.23^{+.22}_{-.23})$ & $1.13^{+.32}_{-.27}$ \nl
\enddata
\tablecomments{Using the full constraint set in Table~3.
Errors, where given, are defined by $\Delta \chi^2=4 \bar{\chi}^2$.}
\end{deluxetable}

\end{document}